\documentclass[a4paper,12pt]{article} 
\usepackage{amssymb}
\usepackage{amscd} 
\usepackage{enumerate}  
\usepackage{latexsym}

\usepackage{theorem}
\theoremstyle{plain}
\theorembodyfont{\itshape}
\newtheorem{theorem}{Theorem}[section]

\newtheorem{lemma}[theorem]{Lemma}

\theorembodyfont{\rmfamily}

\theoremstyle{break}
\theorembodyfont{\itshape}
 
\def\dfrac#1#2{{\displaystyle\frac{#1}{#2}}} 
\def\Ai{\mathrm{Ai}}
 
\begin{document} 
\title{Generating Function Associated with the Rational Solutions of the 
Painlev\'e II Equation} 
\author{Katsunori Iwasaki, Kenji Kajiwara and Toshiya Nakamura \\ \\ 
Graduate School of Mathematics\\
Kyushu University\\
6-10-1 Hakozaki, Higashi-ku, Fukuoka 812-8581 Japan} 
\date{}
\maketitle 
\begin{abstract} 
We consider the Hankel determinant representation for the rational
 solutions of the Painlev\'e II equation.
We give an explicit formula for the generating function of the entries
in terms of logarithmic derivative of the Airy function, which by itself is a
 particular solution of the Painlev\'e II equation.
%All the rational solutions of the Painlev\'e II equation can be 
%represented in terms of Hankel determinants with entries {}from an 
%infinite sequence of certain special polynomials.       
%Hence it is very interesting to consider a generating function for 
%those polynomials.     
%In this paper we find out an explicit formula for the generating 
%function in terms of the Airy function.    
\end{abstract}
\section{Introduction}  \label{sec:introduction} 
In this paper we introduce a generating function associated 
with the rational solutions of the Painlev\'e II equation and 
characterize it explicitly in terms of the Airy function.   
Here the Painlev\'e II equation (P$_{\rm II}$) is a second order nonlinear 
ordinary differential equation with a parameter $\alpha$,  
\begin{equation}
\label{p2}
\dfrac{d^2 v}{dx^2} = 2 v^3 - 4 x v + 4 \alpha.     
\end{equation}
\par
This equation has exactly one rational solution for $\alpha$ being 
an arbitrary integer \cite{A} and has no rational solution if 
$\alpha$ is not an integer \cite{M,O1,UW}.  
It admits a B\"acklund transformation 
$(v, \alpha) \mapsto (-v,-\alpha)$,  
which clearly maps rational solutions to rational solutions. 
Moreover the unique rational solution for $\alpha = 0$ is the 
trivial solution $v \equiv 0$.    
Hence $\alpha$ may be restricted to positive 
integers without loss of generality. 
\par  
According to \cite{KO}, any rational solution admits two types 
of determinant representation; one is the Jacobi-Trudi type 
and the other is the Hankel type, the latter being 
described as follows: For each positive integer $N+1$, the 
unique rational solution for $\alpha = N+1$ is given by    
\[  
v = \dfrac{d}{dx} \log \dfrac{\sigma_{N+1}}{\sigma_N}
\]
where $\sigma_N$ is the Hankel determinant 
\[
\sigma_N = \left|
\begin{array}{cccc}
a_0 & a_1 & \cdots & a_{N-1} \\
a_1 & a_2 & \cdots & a_N \\
\vdots & \vdots & \ddots & \vdots \\
a_{N-1} & a_N & \cdots & a_{2N-2} 
\end{array}
\right|  
\]
with $a_n = a_n(x)$ being polynomials defined by the 
recurrence relation 
\begin{equation} \label{eqn:recurrence} 
\begin{array}{rcl} 
    a_0 &=& x, \qquad a_1 = 1,  \\
\vspace{-0.3cm} & &             \\ 
a_{n+1} &=& \dfrac{d a_n}{dx} + \displaystyle 
\sum_{k=0}^{n-1} a_k \, a_{n-1-k}.  
\end{array} 
\end{equation} 
\par 
On one hand, the determinant representation of the 
Jacobi-Trudi type implies that the rational solutions 
can be described in terms of certain specializations 
of Schur polynomials \cite{KO}. 
What, on the other hand, does the determinant 
representation of the Hankel type mean\,? 
Or more simply we may ask: What is the sequence $a_n$\,? 
\par
As will be seen in this paper, an answer to the above 
question is very intriguing!    
The generating function for the sequence $a_n$ is 
essentially the Airy function. 
More precisely we have the following:       
\begin{theorem} \label{thm:main} 
Let $\theta(x,t)$ be an entire function of two variables 
defined by 
\begin{equation}  \label{eqn:theta}
\theta(x,t) = \exp\left(2 t^3/3 \right)\, \Ai(t^2-x)  
\end{equation}  
where $\Ai(z)$ is the Airy function. 
Then there exists an asymptotic expansion 
\begin{equation} \label{eqn:main}
\dfrac{\partial}{\partial t} \log \theta(x,t) \sim 
\sum_{n=0}^{\infty} a_n(x) \, (-2t)^{-n}  
\end{equation} 
as $t \to \infty$ in any proper subsector of the sector 
$|\arg t|< \pi/2$.  
\end{theorem}  
\par
The occurrence of the Airy function in Theorem \ref{thm:main}  
is quite suggestive to those who know that P$_{\rm II}$ (\ref{p2}) has a particular solution  
\[
v = \dfrac{d}{dx} \log \Ai\left(2^{1/3} x \right), 
\qquad \alpha = 1/2.  
\]
As is well known, P$_{\rm II}$ (\ref{p2}) admits 
(exactly) two classes of classical solutions, namely, 
the class of rational solutions and that of Airy function 
solutions \cite{UW}.  
Theorem \ref{thm:main} might then be interpreted as asserting 
that all members of the former class are generated by 
a particular member of the latter; see \S\ref{sec:concluding} 
for further discussions.    
\section{Riccati Equation} \label{sec:riccati} 
We will find out a differential equation satisfied by the  
generating function 
\begin{equation} \label{eqn:generating} 
F(x,t) = \sum_{n=0}^{\infty} a_n(x) \, (-2t)^{-n},  
\end{equation}
which is thought of as a formal power series of $t^{-1}$ 
with polynomial coefficients in $x$. 
To this end we have to deduce some other recurrence relations 
for $a_n$ {}from the original one (\ref{eqn:recurrence}) together 
with the data on the first three terms   
\begin{equation}
a_0 = x, \qquad a_1 = 1, \qquad a_2 = x^2.  \label{eqn:initial}   
\end{equation} 
\begin{lemma} \label{lem:recurrence} 
The infinite sequence $a_n$ satisfies recurrence relations 
\begin{eqnarray} 
\dfrac{d a_{n+1}}{dx} &=& 2n a_{n-1},  \label{eqn:recurrence1} \\
a_{n+1} &=& 2 (n-1) a_{n-2} + \displaystyle 
\sum_{k=0}^{n-1} a_k \, a_{n-k-1}.  \label{eqn:recurrence2}
\end{eqnarray} 
\end{lemma}  
{\it Proof}.  
Recurrence relation (\ref{eqn:recurrence1}) was already 
mentioned in \cite[(58)]{KO}, but a proof is included 
here for the sake of completeness. 
The proof proceeds by induction on $n$. 
Assume that (\ref{eqn:recurrence1}) holds for $1,\dots,n$. 
Recurrence relation (\ref{eqn:recurrence}) and induction 
hypothesis lead to 
\[
a_{n+2} = a'_{n+1} + \sum_{k=0}^n a_k \, a_{n-k} 
= 2 n a_{n-1} + \sum_{k=0}^n a_k \, a_{n-k} 
\]
Differentiating both sides and using the initial condition 
(\ref{eqn:initial}), induction hypothesis and the recurrence 
relation (\ref{eqn:recurrence}), one finds    
\begin{eqnarray*}
a'_{n+2}
&=& 2n a'_{n-1} + 2 a_n 
+ \displaystyle \sum_{k=2}^n a'_k \, a_{n-k} 
+ \displaystyle \sum_{k=0}^{n-2} a_k \, a'_{n-k} \\
&=& 2n a'_{n-1} + 2 a_n 
+ \displaystyle \sum_{k=2}^n 2(k-1) \, a_{k-2} \, a_{n-k} \\
& & \phantom{2n a'_{n-1} + 2 a_n}  
+ \displaystyle \sum_{k=0}^{n-2} a_k \, 
\cdot\, 2(n-k-1) \, a_{n-k-2} \\ 
&=& 2 a_n + 2n a'_{n-1} 
+ \displaystyle \sum_{k=0}^{n-2} 2n \, a_k \, a_{n-k-2} \\
&=& 2 a_n + 2n a_n \\
&=& 2(n+1) \, a_n.    
\end{eqnarray*} 
Hence (\ref{eqn:recurrence1}) holds for $n+1$ and the induction 
is complete. 
Recurrence relation (\ref{eqn:recurrence2}) is obtained by 
substituting (\ref{eqn:recurrence1}) into (\ref{eqn:recurrence}). 
\hfill $\Box$  
\par\smallskip  
Lemma \ref{lem:recurrence} leads to a differential equation of 
the Riccati type.  
\begin{lemma} \label{lem:riccati}  
The generating function $F(x,t)$ satisfies a differential 
equation 
\begin{equation} \label{eqn:riccati} 
t \dfrac{\partial f}{\partial t} + t f^2 
-\left(4 t^3 + 1 \right) f + 4 x t^3 - 2 t^2 = 0.  
\end{equation} 
\end{lemma}
{\it Proof}. 
{}From (\ref{eqn:generating}) one has 
\begin{eqnarray*}
t F^2 
&=& t \displaystyle \sum_{n=0}^{\infty} 
\left(\displaystyle \sum_{k=0}^n a_k \, a_{n-k} \right) \, (-2t)^{-n} \\
&=& t x^2 + t \displaystyle \sum_{n=1}^{\infty} 
\left(\displaystyle \sum_{k=0}^n a_k \, a_{n-k} \right) \, (-2t)^{-n} \\
&=& t x^2 + t \displaystyle \sum_{n=1}^{\infty}  
\left(a_{n+2} - 2n \, a_{n-1} \right) \, (-2t)^{-n}  \\
&=& t x^2 + 4 t^3 \displaystyle \sum_{n=3}^{\infty} a_n \, (-2t)^{-n} 
+ \displaystyle \sum_{n=0}^{\infty} (n+1)\, a_n (-2t)^{-n} \\
&=& t x^2 + 4 t^3 \left\{ F - x - (-2t)^{-1} - x^2 (-2t)^{-2} \right\}  
+ \left(- t \dfrac{\partial}{\partial t} + 1 \right)\, F \\
&=& - t \dfrac{\partial F}{\partial t} + 
\left(4 t^3 + 1 \right)\, F - 4 x t^3 + 2 t^2     
\end{eqnarray*}
where (\ref{eqn:initial}) was used in the second and fifth equalities 
and (\ref{eqn:recurrence2}) in the third equality.   
Hence $F$ satisfies the differential equation (\ref{eqn:riccati}).  
\hfill $\Box$ 
\section{The Airy Function} \label{sec:airy}  
The Riccati equation (\ref{eqn:riccati}) can be linearized 
in a standard manner.  
\begin{lemma} \label{lem:linear}  
The change of dependent variable  
\begin{equation} \label{eqn:cole}  
f = \dfrac{\partial}{\partial t} 
\log\left\{u \exp\left(2 t^3/3 \right) \right\}   
\end{equation}
transforms the Riccati equation $(\ref{eqn:riccati})$ into 
a linear equation 
\begin{equation} \label{eqn:linear}  
t \dfrac{\partial^2 u}{\partial t^2} 
- \dfrac{\partial u}{\partial t} 
- 4 \left(t^5 - x t^3 \right) u = 0. 
\end{equation} 
Furthermore the change of independent variable 
\begin{equation} \label{eqn:transf}  
z = t^2 - x 
\end{equation}
simplifies $(\ref{eqn:linear})$ into the Airy differential 
equation   
\begin{equation} \label{eqn:airy}   
\dfrac{\partial^2 u}{\partial z^2} -z u = 0.  
\end{equation} 
\end{lemma}
\par
The proof is just by direct calculations.  
\par\smallskip 
It is well known \cite{L,W} that the Airy equation 
(\ref{eqn:airy}) has the formal solutions  
\begin{equation} \label{eqn:formalsol}
U_{\pm}(z) = \dfrac{1}{2 \sqrt{\pi}} 
\exp\left(\pm \dfrac{2z^{3/2}}{3} \right) \, z^{-1/4} \,  
\sum_{n=0}^{\infty} \dfrac{(1/6)_n (5/6)_n}{n!} \, 
\left(\pm \dfrac{4 z^{3/2}}{3} \right)^{-n}  
\end{equation}
and that the Airy function $\Ai(z)$ admits an asymptotic 
representation 
\begin{equation} \label{eqn:airyasymp}
\Ai(z) \sim U_-(z)
\end{equation}
as $z \to \infty$ in any proper subsector of the sector 
$|\arg z| < \pi$. 
\begin{lemma} \label{lem:formal} 
Equation $(\ref{eqn:linear})$ admits a formal solution 
\begin{equation} \label{eqn:formal}   
U(x,t) 
= \dfrac{1}{2 \sqrt{\pi}} 
\exp\left(-\dfrac{2t^3}{3} \right) \, t^{-1/2} \, 
\exp\left\{\dfrac{1}{2}  
\sum_{n=1}^{\infty} \dfrac{a_{n+1}(x)}{n} \, (-2t)^{-n} \right\} 
\end{equation} 
If the branch $z^{1/2} = t (1-x t^{-2})^{1/2} 
= t -(x/2) t^{-1} +\cdots$ is taken for the 
square root of $(\ref{eqn:transf})$, then 
\begin{equation} \label{eqn:airyformal} 
U(x,t) = U_-(z) 
\end{equation}    
\end{lemma} 
{\it Proof}. 
Observe that $u = U(x,t)$ defined by (\ref{eqn:formal}) 
satisfies the equation (\ref{eqn:cole}) with $f = F(x,t)$.  
Since $F(x,t)$ solves the Riccati equation (\ref{eqn:riccati}), 
Lemma \ref{lem:linear} implies that (\ref{eqn:formal}) solves 
the linear equation (\ref{eqn:linear}). 
Under the relation (\ref{eqn:transf}), $U(x,t)$ must be either 
$U_+(z)$ or $U_-(z)$ since it is a formal solution of the Airy 
equation (\ref{eqn:airy}).  
Comparing the exponential factors of (\ref{eqn:formalsol}) 
and (\ref{eqn:formal}), we must choose the minus sign and hence 
have (\ref{eqn:airyformal}). \hfill $\Box$ 
\par\smallskip\noindent   
{\it Proof of Theorem $\ref{thm:main}$}.  
We are now in a position to establish Theorem \ref{thm:main}. 
By (\ref{eqn:airyasymp}) and (\ref{eqn:airyformal}) the 
function $\Ai(t^2-x)$ has an asymptotic representation 
$U(x,t)$ as $t \to \infty$ in any proper subsector of the 
sector $|\arg t| < \pi/2$ (note that under the relation 
(\ref{eqn:transf}) a sector of central angle $\theta$ in 
the $z$ plane corresponds to one of central angle 
$\theta/2$ in the $t$ plane).    
This fact, through the transformation (\ref{eqn:cole}), 
leads to the asymptotic expansion (\ref{eqn:main}). 
\hfill $\Box$   
\section{Concluding Discussions} \label{sec:concluding} 
In this paper we have considered the infinite sequence of 
polynomials that appears in the Hankel determinant 
representation for rational solutions of the Painleve II 
equation. As a result we have constructed a generating function explicitly represented in 
terms of the Airy function, which by itself is a particular 
solution of the Painleve II equation.  
\par
It is natural to ask whether there are similar 
phenomena for other Painlev\'e equations.
It is known that any other Painlev\'e equation,
%(when restricted to special parameters)
except for the first one P${}_{\mathrm{I}}$, 
also admits two classes of classical solutions, namely, the 
class of rational solutions and that of special function 
solutions; Bessel for P${}_{\mathrm{III}}$, Hermite-Weber for 
P${}_{\mathrm{IV}}$, Kummer for P${}_{\mathrm{V}}$ and Gauss for 
P${}_{\mathrm{VI}}$ \cite{O1,O2,O3,O4}.    
It is also known that all the rational solutions are represented   
in terms of Hankel determinants with entries {}from an infinite 
sequence of certain special polynomials \cite{KMNOY}.  
Now a natural question is whether this infinite sequence has a 
generating function expressible in a closed form 
by using the respective special functions. Moreover, how about 
generic (transcendental) solutions? Do they admit similar phenomena?
Further, how about the discrete cases? 
These points, together with the mechanism behind these strange phenomena,
are yet to be explored in the future.              
\end{document}